\documentclass[a4paper,11pt]{article}
\usepackage{pos}

\newcommand{\be}{\begin{equation}}
\newcommand{\ee}{\end{equation}}
\newcommand{\eq}[1]{Eq.~(\ref{#1})}
\newcommand{\fig}[1]{Fig.~\ref{#1}}

\def\tr{{\rm tr}\,}

\def\Nf{N_\text{f}}

\def\gbar{\bar g}

\newcommand{\gGF}{\overline{g}_{\rm GF}}

\newcommand{\beann}{ \begin{eqnarray*}}
\newcommand{\eeann}{ \end{eqnarray*}  }

\newcommand{\bea}{\begin{eqnarray}}
\newcommand{\eea}{\end{eqnarray}}
\newcommand{\bean}{\begin{eqnarray*}}
\newcommand{\ean}{\end{eqnarray*}}

\def\permille{\ensuremath{{}^\text{o}\mkern-5mu/\mkern-3mu_\text{oo}}}

\def\tr{{\rm tr}\,}

\def\Nf{N_{\rm{f}}}

\def\gbar{\bar g}

\title{Results for $\alpha_s$ from the decoupling strategy\\\href{https://www-zeuthen.desy.de/alpha/}{ALPHA collaboration}}
\ShortTitle{Results for $\alpha_s$ from the decoupling strategy}

\author[a]{Mattia~Dalla~Brida}
\author*[b]{Roman H{\"o}llwieser} 
\author[b]{Francesco Knechtli}
\author[b]{Tomasz Korzec}
\author[c]{Alessandro~Nada}
\author[d]{Alberto~Ramos} 
\author*[e]{Stefan~Sint}
\author[c,f]{Rainer~Sommer}

\affiliation[a]{Theoretical Physics Department, CERN, 1211 Geneva 23, Switzerland}
\affiliation[b]{Deptartment of Physics, University of Wuppertal, Gau{\ss}strasse 20, 42119 Germany}
\affiliation[c]{John von Neumann Institute for Computing (NIC), DESY, Platanenallee~6, 15738~Zeuthen, Germany}
\affiliation[d]{Instituto de F\'isica Corpuscular (IFIC), CSIC-Universitat de Valencia, 46071, Valencia, Spain}
\affiliation[e]{School of Mathematics and Hamilton Mathematics Institute, Trinity College Dublin, Dublin 2, Ireland}
\affiliation[f]{Institut~f\"ur~Physik, Humboldt-Universit\"at~zu~Berlin, Newtonstr.~15, 12489~Berlin, Germany}

\emailAdd{hoellwieser@uni-wuppertal.de}
\emailAdd{sint@maths.tcd.ie}

\abstract{We present analysis details and new results for the strong coupling $\alpha_s(m_Z)$, determined by the decoupling strategy. 
We measure a massive gradient flow (GF) coupling defined in finite volume with Schr\"odinger functional (SF) boundary conditions in a theory with $\Nf=3$ degenerate heavy quarks of mass $M$.
The massive couplings are matched to effective couplings in pure gauge. Using the running in the pure gauge theory and the perturbative relation of the Lambda parameters, the Lambda parameter of the three flavor theory is obtained by an extrapolation to infinite M. Our final result is compatible both with the FLAG average and with the previous ALPHA result, albeit with a slightly smaller, yet still statistics dominated, error. This constitutes a non-trivial check, as the decoupling strategy is conceptually very different from the 3-flavor QCD step-scaling method, and so are most of its systematic errors. These include the uncertainties of the decoupling and continuum limits, which we discuss in some detail. Furthermore, by relying on decoupling once again, we could estimate the small $O(a)$ and $O(1/M)$ contaminations to the massive GF coupling stemming from the SF boundaries by means of pure gauge simulations.\\

\begin{flushright}
CERN-TH-2021-207, WUB/21-03
\end{flushright}
}

\FullConference{%
 The 38th International Symposium on Lattice Field Theory, LATTICE2021
  26th-30th July, 2021
  Zoom/Gather@Massachusetts Institute of Technology
}

\begin{document}
\maketitle

\section{Introduction}

The determination of the running coupling $\alpha_s=\overline g^2/(4\pi)$ of the strong force yields~\cite{Aoki:2021kgd,Ayala:2020odx, Bazavov:2019qoo, Cali:2020hrj,Aoki:2019cca,Maltman:2008bx,Aoki:2009tf,McNeile:2010ji,Chakraborty:2014aca,Bazavov:2014soa,Nakayama:2016atf,Bruno:2017gxd}  with most precise results from lattice QCD based on finite volume renormalization schemes.
In~\cite{DallaBrida:2019mqg} we implement a new strategy to extract $\alpha_s$ from lattice
QCD simulations based on the decoupling relation for a massive coupling
\begin{equation}\label{eq:dec}
\overline g^2_{\Nf=3}(\mu, M) = \overline g^2_{\Nf=0}(\mu) + O((\Lambda/M)^2, (\mu/M)^2)\, .
\end{equation}
Here $\overline g^2_{\Nf=3}(\mu, M)$ is  a renormalized coupling in QCD  with $\Nf=3$ massive quarks
of mass $M$\footnote{We follow the notation of \cite{Athenodorou:2018wpk} and
denote by $M$ the renormalization group invariant (RGI) quark mass, and by $\Lambda$ the Lambda-parameter of QCD in the $\overline{\text{MS}}$ scheme.} and $\overline g^2_{\Nf=0}(\mu)$ is the coupling in the pure gauge theory. The renormalization scale,
$\mu$, is the same in both theories.
The result of Ref.~\cite{Bruno:2017gxd} is based on the  non-perturbative running of the coupling from low to high energies. 
Eq.~\eqref{eq:dec} defers this computation
to the pure gauge theory,
where very high precision can be achieved~\cite{DallaBrida:2019wur,Nada:2020jay}.
We computed a finite volume coupling in a setting with $\Nf=3$ mass-degenerate heavy quarks for
values of the quark mass ranging from charm to above the bottom and already provided
a proof of principle that Eq.~\eqref{eq:dec} can be used to extract $\alpha_s$
\cite{DallaBrida:2019mqg}. Here we present our latest results,  
confirming the world average of $\alpha_s$ with another independent method, and a good chance to further reduce its uncertainty.

\section{Strategy}

\begin{figure}[h]
 \centering
\includegraphics[width=0.495\textwidth]{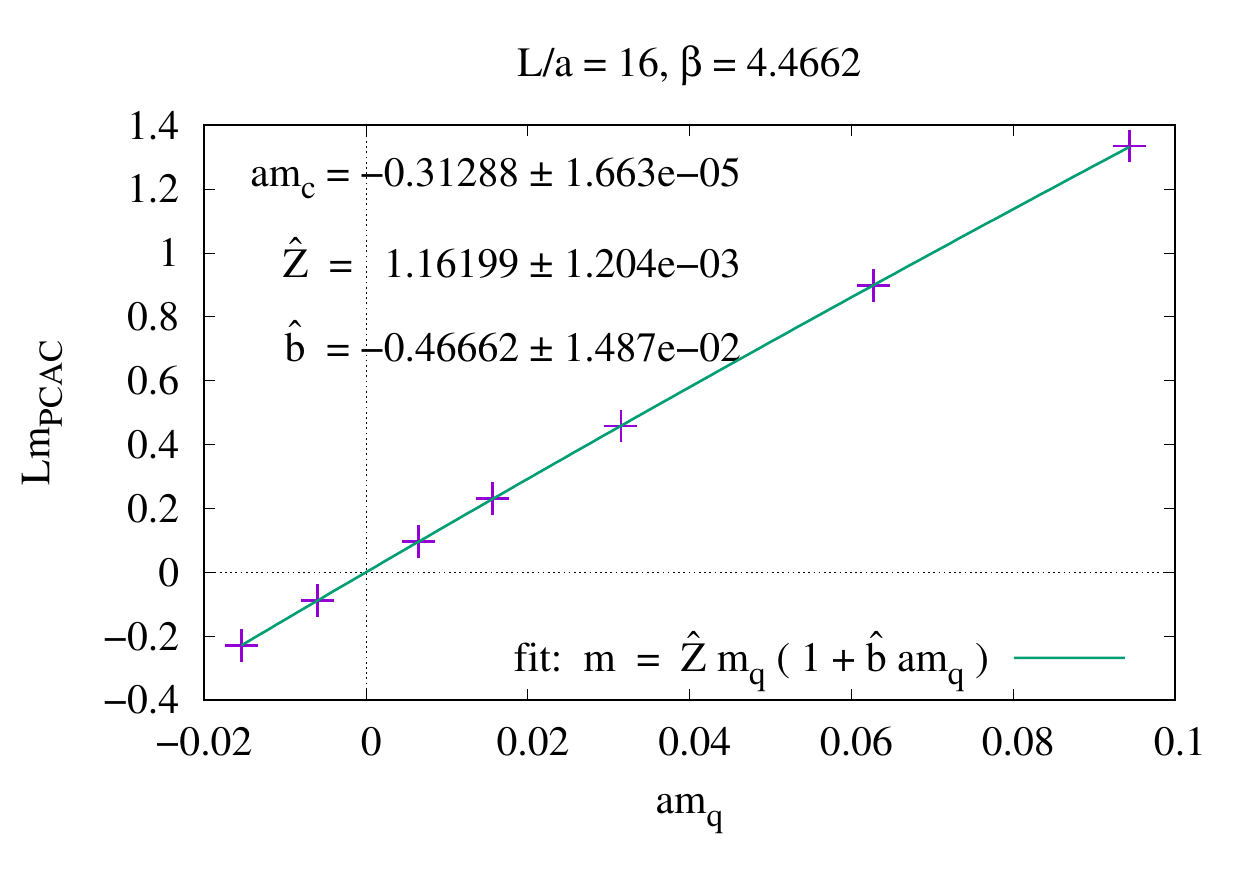}
\includegraphics[width=0.495\linewidth]{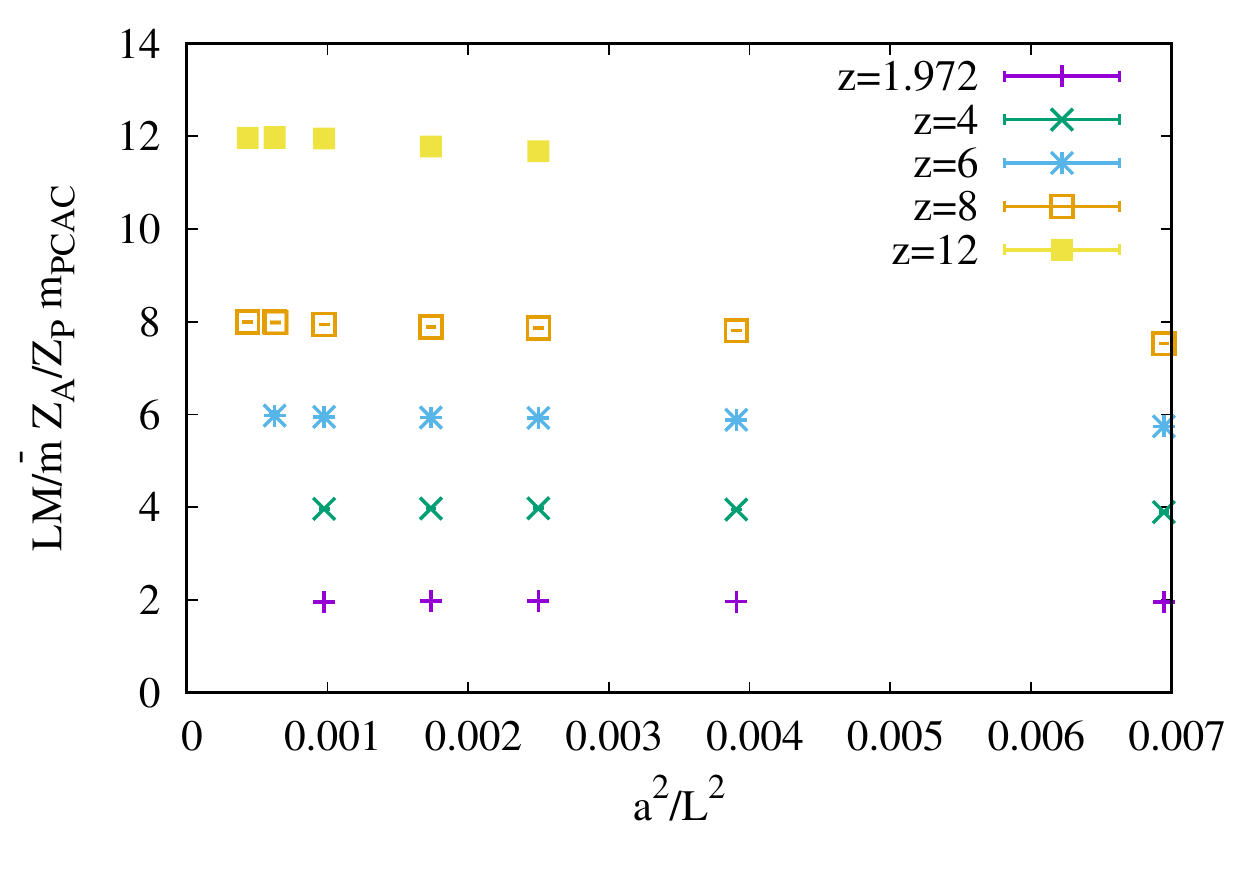} \caption{(left) Bare mass $m_\mathrm{PCAC}$ as a function 
 of the bare subtracted mass $m_\mathrm{q}=1/2\kappa -1/2\kappa_c$ for various values of 
 $\kappa$ and $g_0^2$ fixed as explained. The fit
 determines $\kappa_c,\hat Z, \hat b$ for the resolution $L/a=16$ and with Schr\"odinger Functional boundary conditions and $T=L$. (right) Renormalized PCAC masses of the massive simulations, giving an overview of simulated resolutions $L/a$ and quark masses $M$, also providing a non-trivial check of our simulation parameters as simulated (intended) $z$ values correspond to measured and renormalized $LM$ values in the continuum limit.
 }
 \label{fig:massless}
\end{figure}

Decoupling~\cite{Appelquist:1974tg, Weinberg:1980wa} applies to dimensionless, renormalized, low-energy, quantities which include suitably defined couplings at low renormalization scales. Eq.\eqref{eq:dec} holds when the 
two theories are matched, i.e. that the $\Lambda$-parameter of the $\Nf=0$ theory is chosen
such that
\begin{equation}
\label{eq:lammatch}
   \Lambda^{(0)}_{\overline{\rm MS}} = \Lambda^{(3)}_{\overline{\rm MS}}\ P_{0,3}(M / \Lambda^{(3)}_{\overline{\rm MS}})\, .
\end{equation}
$P_{0,3}$ is the matching factor between the two theories, which, if $M$ is large enough, can be computed very 
accurately in perturbation theory~\cite{Bernreuther:1981sg,Chetyrkin:2005ia,Schroder:2005hy,Kniehl:2006bg,Grozin:2011nk,Gerlach:2018hen,Athenodorou:2018wpk}.
Equations {\eqref{eq:dec}, \eqref{eq:lammatch}} can be exploited to determine the three flavor $\Lambda$-parameter following these
steps:
\begin{itemize}
   \item Choose a low energy renormalization scale $\mu$, that is known in physical units (MeV) in the
	 three flavor theory.
   \item Determine a massive coupling $\overline g^2=\overline g^2_{\Nf=3}(\mu, M)$ on lattices with different lattice 
	 spacing $a$, and take the continuum limit.
   \item Determine the non-perturbative $\beta$-function of the  coupling in the $\Nf=0$ theory and compute the
	 $\Lambda$-parameter in units of $\mu$ \cite{DallaBrida:2019wur}
	 \begin{equation}\label{eq:Lambda}
		 \frac{\Lambda^{(0)}}{\mu} = (b_0 \overline g^2)^{-b_1/(2b_0^2)}\ e^{-1/(2b_0\overline g^2)}\ \exp\left\{-\int\limits_0^{\overline g} 
                   \left[\frac{1}{\beta(x)}+\frac{1}{b_0x^3}-\frac{b_1}{b_0^2 x} \right]\,dx\right\} \, .
         \end{equation}
	 The $\Lambda$-parameter in the $\overline{\rm MS}$  scheme is then given {\em exactly} by a 1-loop relation. For more details on the exact procedure see~\cite{DallaBrida:2019mqg}.
   \item Obtain the $\Nf=3$ $\Lambda$-parameter in physical units as
         \begin{equation}\label{eq:master}
		 \Lambda^{(3)}_{\overline{\rm MS}} = \mu \times \frac{\Lambda^{(0)}_{\overline{\rm MS}}}{\mu} \times \frac{1}{P_{0,3}(M / \Lambda^{(3)}_{\overline{\rm MS}})} + O(M^{-2})\, .
	 \end{equation}
\end{itemize}
Finally, the result can be translated to the commonly used coupling constant $\alpha_s^{(5)}(M_Z)$, relying on the use of perturbation theory in the $\overline{\rm MS}$ scheme at the charm and bottom mass thresholds.
In~\cite{DallaBrida:2019mqg} we have shown, that this strategy is viable and
able to reduce the uncertainty of the strong coupling. Several of the above steps were already carried
out within different projects, {\it e.g.}, we know that in the three flavor theory
$\gGF^2(\mu, M=0)=3.95$ implies $\mu = 789(15)$ MeV~\cite{DallaBrida:2016kgh,Bruno:2017gxd}, 
where $\gGF$ denotes the gradient-flow coupling, 
that runs with the box-size of the system $\mu=1/L$~\cite{Fritzsch:2013je}. 
We denote this particular choice of renormalization scale by $\mu_{\rm dec}$ from here on. It is the 
low-energy scale (low in the sense that $\mu \ll M$) at which decoupling in the form of~eq.~(\ref{eq:dec})
is applied.
Another important ingredient that has already been worked out, is the $\beta$-function of
the $\Nf=0$ gradient-flow coupling. It has been constructed non-perturbatively to a very high precision,
such that Eq.~(\ref{eq:Lambda}) can be evaluated for a large range of couplings~\cite{DallaBrida:2019wur}.
The other key ingredient is a precise determination of the {\em massive} coupling at scale $\mu$
for various $M$. These simulations have to 
follow so-called lines of constant physics which give 
the bare parameters of the discretized theory 
such that 
\begin{equation}
	\gGF^2(\mu_\mathrm{dec},0)=3.95\,,\quad M/\mu_\mathrm{dec} \equiv z \in \{2,4,6,8,12\}\label{eq:z}
\end{equation}
for various resolutions $a/L=a\mu_\mathrm{dec}$.
In a B-physics project of our collaboration~\cite{Fritzsch:2019cf}, bare couplings, $\tilde g_0^2$ and  hopping parameters 
$\kappa=\kappa_\mathrm{crit}$ have been tuned such that the
condition for the massless coupling is satisfied 
within less than $4\permille$ and such that indeed
the quarks are massless to high accuracy. The chosen resolutions are $L/a=12,16,20,24,32$ and $40$, where the parameters of $L/a=40$ can be inferred from \cite{DallaBrida:2016kgh}. For the massive coupling, fixed values of $z=ML$ determine $aM$ and therefore the hopping parameter $\kappa$. Their relation is provided by the  following renormalization 
\begin{equation}
   M = \frac{M}{\overline m(\mu)} \frac{Z_A(\tilde g_0)}{Z_P(\tilde g_0,\mu)} m_{\rm PCAC} \left(1 + (b_A-b_P)a m_q \right)\, ,
   \quad 
   m_{\rm PCAC} = \hat Z(\tilde g_0) m_q (1 + \hat b\, am_q)\, ,
\end{equation}
where $am_q = 1/(2\kappa) - 1/(2\kappa_{\rm crit})$ is the bare subtracted quark mass.
All parameters in the relation between PCAC mass and RGI quark mass
are known from~\cite{Campos:2018ahf} and $Z_A$ from~\cite{DallaBrida:2018tpn}. We have
carried out massless MC simulations to determine $\hat Z$ and $\hat b$ with an example depicted in~\fig{fig:massless} on the left.
The knowledge of these parameters allows massive MC simulations of 
$L/a=12,16,20,24,32,40$ with $T=2L$ and for the $z$-values in~\eq{eq:z} we have $M\approx1.6\ldots9.5$ GeV, see the right plot of~\fig{fig:massless} for an overview of massive simulations to be discussed below. The fine resolutions are crucial to fully control the continuum limit $a/L\to0$ of 
\begin{equation}
	\gbar^2_{\Nf=3}(\mu, z)\quad\text{for}\quad  \gbar^2_{\Nf=3}(\mu, 0)=3.95,\; \mu=1/L,\;z=LM\,,
\end{equation}
which gives values for the massive coupling in the 
continuum limit. The second limit that needs to be controlled is $M\to\infty$ in eq.~\eqref{eq:master}. 

Before we discuss how we obtain numerical control over the double limit 
$\lim_{z\to\infty} \lim_{a/L\to0}$, we need to explain some details on the simulations, in particular the definition of the non-perturbative coupling, associated systematic effects and how we control them. 

\section{Simulations and Analysis}

For the Monte-Carlo simulations we use the open-source (GPL v2) {\tt openQCD} package\footnote{\url{http://luscher.web.cern.ch/luscher/openQCD/}}~\cite{Luscher:2012av} in plain C with MPI parallelization. The software has been successfully used in various large-scale projects and we use it in its version {\tt openQCD-1.6} with additional implementation of
\begin{itemize}
\item the correct Schr\"odinger Functional boundary conditions for the Symanzik improved gauge action with SF boundary conditions precisely as in~\cite{DallaBrida:2016kgh},
\item on-the-fly measurements of gradient-flow observables using the Zeuthen flow~\cite{Ramos:2015baa}, including measurements of the gradient flow coupling and the topological charge,
\item on-the-fly measurements of Schr\"odinger Functional correlators, needed for the determination of the PCAC mass.
\end{itemize}
All simulations in this project use the L\"uscher-Weisz improved gauge action with O($a$) improvement for $\Nf=3$ quarks tuned at and around zero quark mass in the first respectively quark masses $M\approx1.6\ldots9.5$GeV in the second set of simulation runs.

We use a gradient flow coupling in a finite volume 
$T\times L^3$ with Schr\"odinger functional (SF) boundary conditions \cite{DallaBrida:2019mqg}. In this setting the formally leading corrections to decoupling are not $1/M^2$ but they are $1/M$.
In the low energy effective field theory the $1/M$ term
originates from only one operator, $\tr F_{0k} F_{0k}$ located at the two time-boundaries of the SF manifold. The exactly same term is responsible 
for $O(a)$ terms of the pure gauge SF. Its renormalization group improved perturbative 
expansion  has recently been discussed \cite{Husung:2019ytz}. In complete 
analogy we are able to treat the $1/M$ term and show 
that it is very small.  Its smallness is due to a combination of 1) the smallness of the coefficient in the effective theory, which follows from \cite{Sint:1995ch}, 2) the vanishing of the anomalous dimension of  $\tr F_{0k} F_{0k}$ at the boundary \cite{Husung:2019ytz} and 3) our choice $T=2L$. 
The latter was a precaution that we took in \cite{DallaBrida:2019mqg}. We are presently working out the coefficient of $\tr F_{0k} F_{0k}$ to next to leading order in perturbation theory in $\gbar^2_{\overline{\rm MS}}(m_\star)$, where $\overline{m}_{\overline{\rm MS}}(m_\star)=m_\star$. The effect of the boundary operator can then be determined in the pure gauge theory.  In summary, while $1/M$ terms are there, they can be estimated well and are negligible. Also $O(a)$ boundary lattice artifacts are suppressed by the choice $T=2L$ and are very small due to the implemented one-loop boundary O($a$) improvement.

The basis for the analysis of the continuum and decoupling limits is determined via first applying Symanzik EFT~\cite{Symanzik:1981hc} and then performing a heavy quark mass expansion of that continuum EFT. The first step tells us that 
the only $\sim a^2$ cutoff effects accompanied by positive powers of the quark mass are of the form $a^2 M^2$, once O($a$) improvement is done. The second step yields a series in
powers of $1/M^2$ of all terms in the Symanzik EFT, when we make the usual assumption that also the second level EFT is described by a local effective Lagrangian. Taking only the
leading corrections, this argumentation yields 
\begin{equation}
\bar g^2(z_i) = c_i + p_1\,[\alpha_s(a^{-1})]^{\hat\Gamma_1}\,(a/L)^2 + p_2\,[\alpha_s(a^{-1})]^{\hat\Gamma_2}\,(aM)^2,\label{eq:fit}
\end{equation}
as a fit function for performing the continuum limit. The presence of
log-corrections of the form $[\alpha_s(a^{-1})]^{\hat\Gamma_i}$
is due to the anomalous dimensions of the operators in the EFTs. There is partial knowledge on them from
 Husung et al.~\cite{Husung:2019ytz,Husung:2021mfl,Husung:2021geh},
 but it is not yet complete. We will vary the $\Gamma_i$ to an extent suggested by \cite{Husung:2019ytz,Husung:2021mfl,Husung:2021geh}, being aware that this is not the end of the story.

\begin{figure*}[h!]
  \centering
  \includegraphics[width=0.495\textwidth]{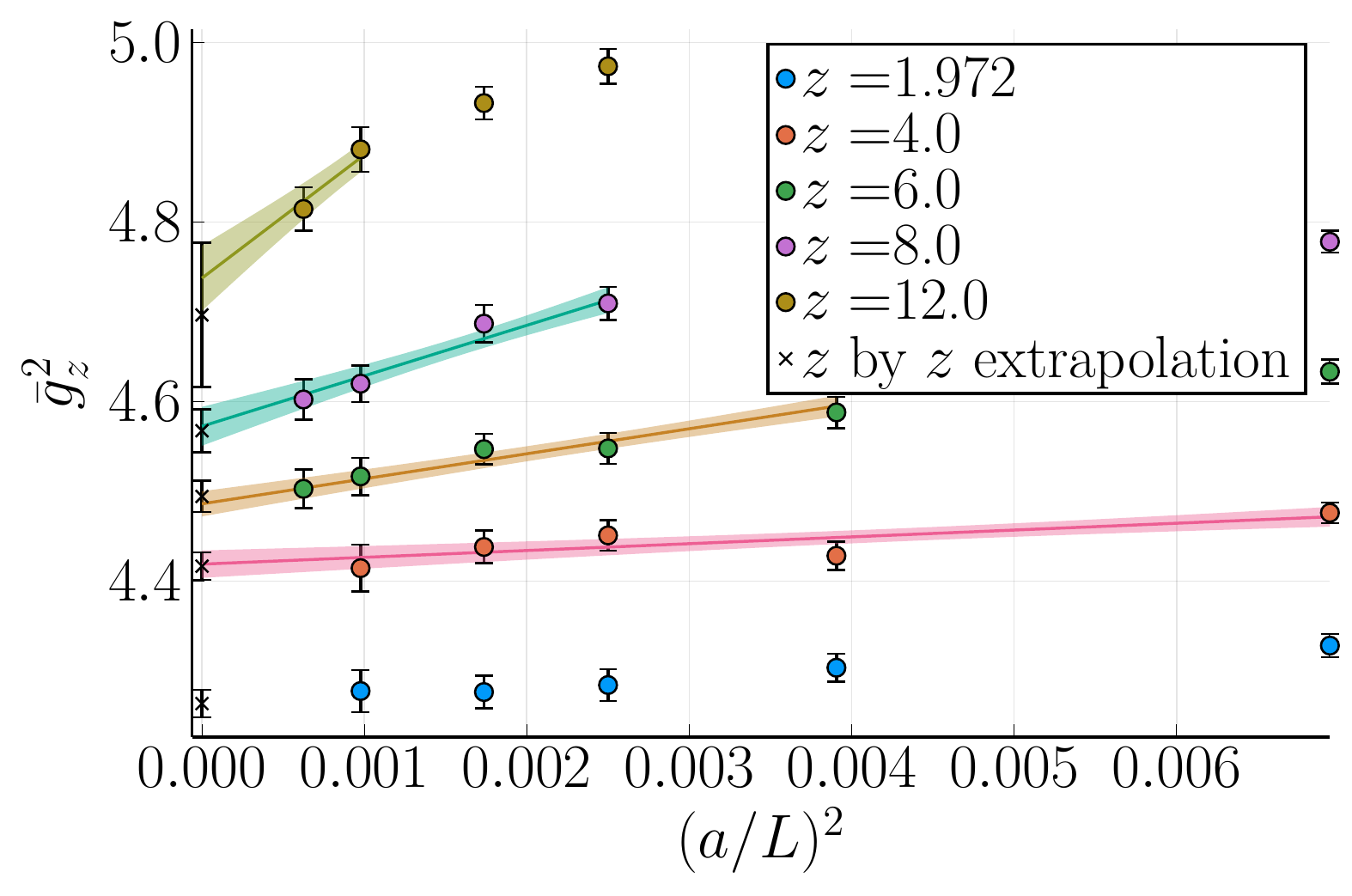} 
  \includegraphics[width=0.495\textwidth]{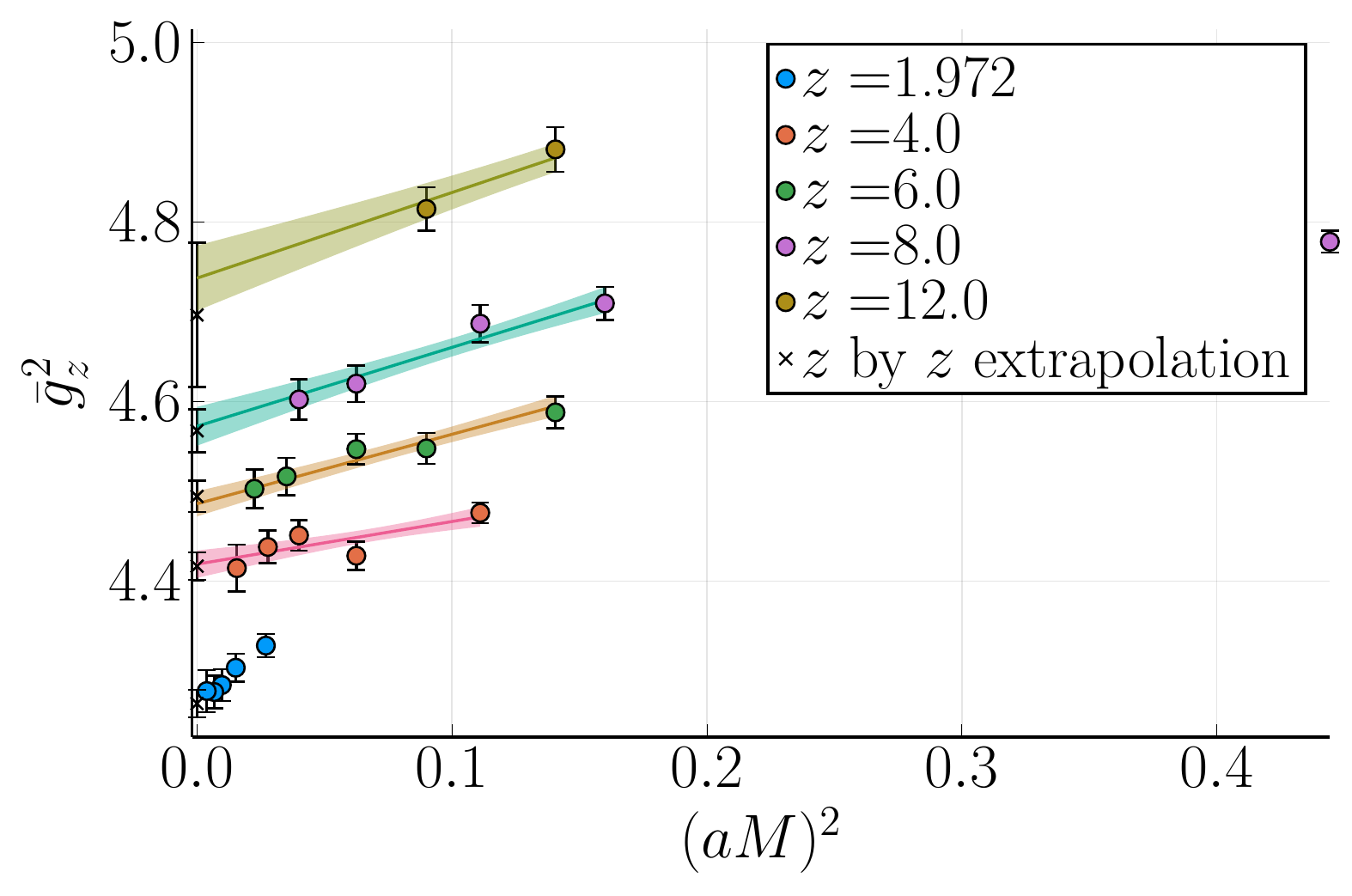} 
  \caption{The gradient flow couplings $\gbar^2_z=\gbar^2(\mu,z\mu)$ of our massive simulations for $z=2,4,6,8,12$ from bottom to top versus the leading discretization effects $(a/L)^2$ (left) and $(aM)^2$ (right), together with the combined, correlated, linear fit in~\eq{eq:fit}, taking into account only data points with $z\geq4$ and $aM\leq0.4$. Note, the crosses on $a=0$ axes stem from individual $z$ extrapolations, not the combined fit.}
  \label{fig:cont}
\end{figure*}

The combined, linear fit of our data using~\eq{eq:fit} is shown in~\fig{fig:cont}. In order to get a good quality of the fit (min. $\chi^2$), we only take data points with $z\geq4$ and $aM\leq0.4$ into account. We see in the right plot of~\fig{fig:cont} that the $z=2$ data shows a very different slope in $(aM)^2$ compared to the other data sets. Further, varying the exponents $\hat\Gamma_i$ in the range $[-1\ldots1]$ gives a systematic error which is negligible in the final result of $\Lambda$. We use $\hat\Gamma_1=\hat\Gamma_2=0$ for our central values.

\begin{figure}[h]
  \centering
  \includegraphics[width=0.505\textwidth]{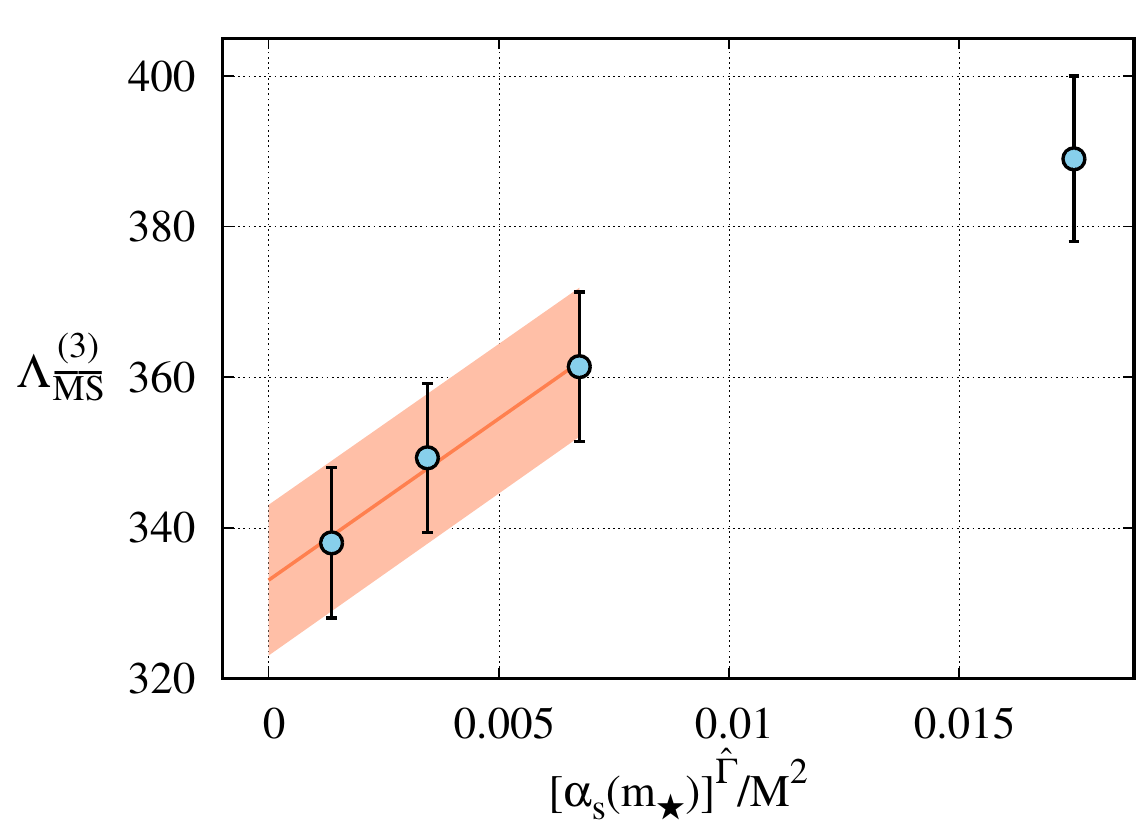} 
  \includegraphics[width=0.489\textwidth]{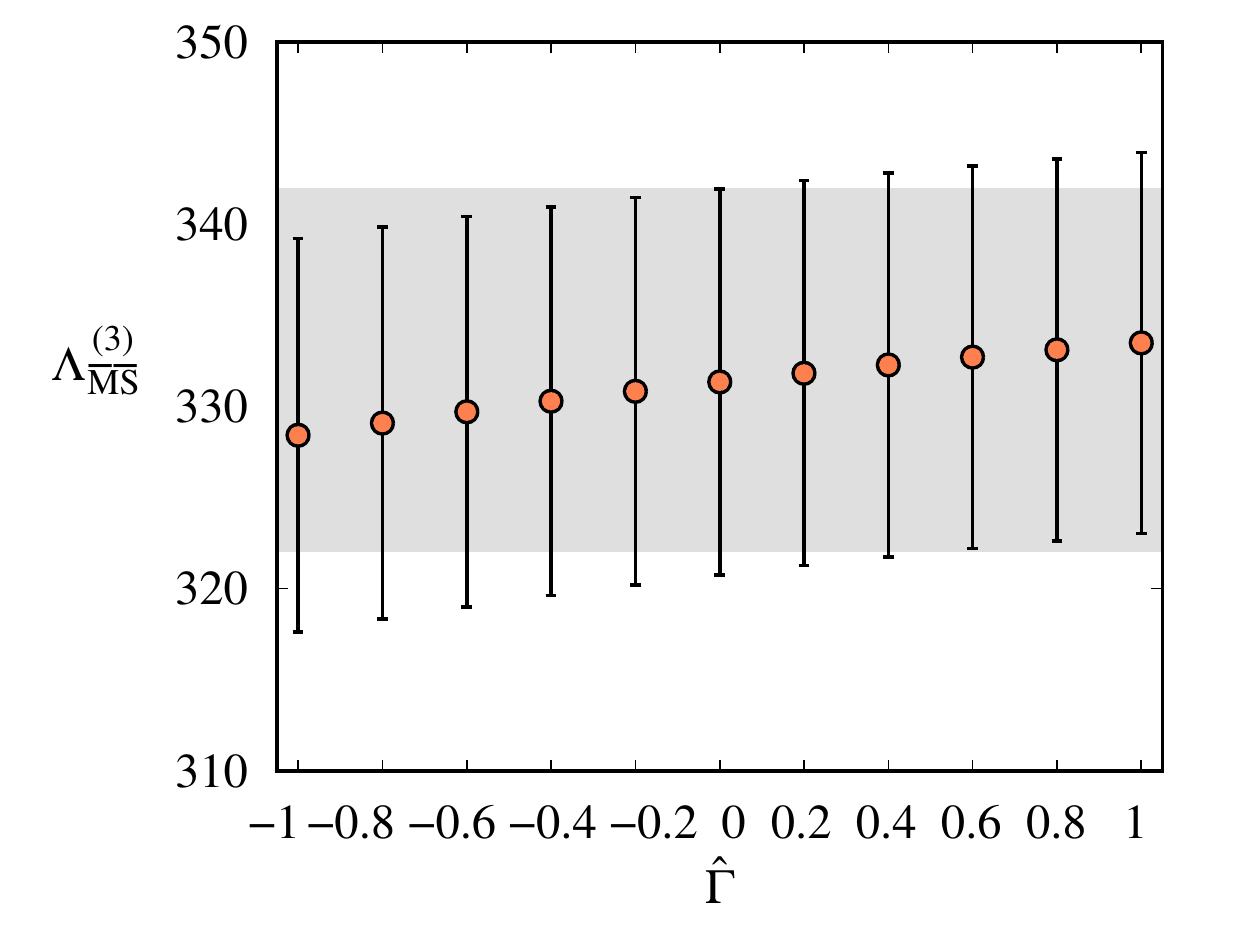} 
  \caption{Values for $\Lambda^{(3)}_{\overline{\rm MS} }$ determined
    from the decoupling relation Eq.~(\ref{eq:master}) and their extrapolation $M\to\infty$ using~\eq{eq:fdec}. In order to get a good quality of the fit (min. $\chi^2$) we only take data points with $z\geq6$ into account (left). The right plot shows different extrapolations for different exponents $\hat\Gamma\in[-1,1]$ of the logarithmic corrections in~\eq{eq:fdec}, with the grey band indicating the preliminary result and error given below.}
  \label{fig:final}
\end{figure}

Using Eqs.~(\ref{eq:Lambda}) and (\ref{eq:master}) we translate our continuum extrapolated couplings $\gbar^2_z=\gbar^2(\mu,z\mu)$ into $\Lambda^{(3)}_{\overline{\rm MS}}$-parameters in physical units. For the decoupling ($M\rightarrow\infty$) extrapolation we find the functional form 
\begin{equation}
\Lambda^{(3)}_{\overline{\rm MS}}(z)=A+\frac{B}{z^2}\,[\alpha_s(m_\star)]^{\hat\Gamma}\qquad\text{with}\quad \overline{m}_{\overline{\rm MS}}(m_\star)=m_\star,\label{eq:fdec}
\end{equation}
where again the fractional exponent $\hat\Gamma$ of the logarithmic correction is not known. In fig.~\ref{fig:final} we fit~\eq{eq:fdec} to the continuum extrapolated $\Lambda^{(3)}_{\overline{\rm MS}}$-parameters, omitting the data point with $z=4$ which is clearly outside $1/z^2$ scaling, and present the extrapolated values for $\hat\Gamma\in[-1,1]$ in the right plot. 

\section{Conclusions and Outlook}

With the present status of the simulations and their continuum ($a/L$ resp. $aM\rightarrow0$) and decoupling ($M\to \infty$) limits, we derive a preliminary value of
\begin{equation*}
	\Lambda^{(3)}_{\overline{\rm MS}}=332(10)(2)\text{ MeV}
\end{equation*}
where we add  a systematic error of 2MeV for the variation with $\hat\Gamma$ in~\eq{eq:fdec}.
The final value is about one standard deviation smaller, but in agreement with the previous result  $\Lambda^{(3)}_{\overline{\rm MS}}=341(12)$MeV~\cite{Bruno:2017gxd}, entirely performed in the $\Nf=3$ theory. We want to stress that the present analysis is a largely independent computation, only the scale $\mu_\text{dec}$ is in common, which contributes an overall $\sim40\%$ to the error squared in the present analysis. The four-loop prediction for $\Lambda^{(5)}_{\overline{\rm MS}}/\Lambda^{(3)}_{\overline{\rm MS}}$ yields
\begin{center}
$\alpha_s(M_Z)=0.1179(7)(1)(1)=0.1179(7)$,
\end{center}
where the first two errors are the translations of the errors $(10)(2)$ for $\Lambda^{(3)}_{\overline{\rm MS}}$ and
the last one is the difference between using perturbation theory with all known orders  and 2 orders less, respectively. In order to reduce the error further, we work on refining the analysis, {\it e.g.}, by fixing the exponents $\hat\Gamma_i$, and including more data points with increased statistics.
Some simulations are still ongoing or yet to be analyzed and we may hope for a reduction of the error of the world average by a factor of two since at the same time, we will use the synergy with other projects of the ALPHA collaboration to reduce uncertainties in other elements which go into the analysis and final result. These are:
\begin{itemize}
	\item The determination of $\mu_\mathrm{dec}$ in physical units \cite{DallaBrida:2019mqg} is based on 1) the scale setting 
	of CLS \cite{Bruno:2016plf,Bruno:2017gxd} and 2) the running of the massless GF coupling between\
	$\mu_\mathrm{dec}$ and $\mu_\mathrm{dec}/4$. These 
	will be improved by:
	\begin{enumerate}
		\item Newer CLS ensembles reach down further in the light quark masses. This will allow for an improved scale setting~\cite{Strassberger:2021}. 
		\item The ALPHA collaboration B-physics project performs extensive simulations in $L_0=1/\mu_\mathrm{dec}$ as well as $L_1=2L_0$, $L_2=4L_0$ volumes. The step scaling function of the massless GF coupling will be determined with higher precision and better resolution than in \cite{DallaBrida:2016kgh} exactly in the range of scales needed here.
	\end{enumerate}
	\item	While the determination of the $\Nf=0$ $\beta$-function in \cite{DallaBrida:2019wur} is very precise, it does contribute a non-negligible amount to the overall uncertainty, which could be further reduced. Note that we are also performing further cross-checks on the determination of the pure gauge theory $\Lambda$-parameter.
\end{itemize}
For more details on our new procedure based on decoupling and an overview of past, present, and future of precision determinations of the QCD coupling from lattice QCD, please see~\cite{DallaBrida:2020pag,DelDebbio:2021ryq,DallaBrida:2022}.

\section*{Acknowledgements}
A.R. acknowledges financial support from the Generalitat Valenciana (genT program CIDEGENT/2019/040) and the Ministerio de Ciencia e Innovación  PID2020-113644GB-I00.
Some authors were supported by the European Union’s Horizon 2020 research and innovation program under the Marie Skłodowska-Curie grant agreement nos. 813942 (ITN EuroPLEx) (S.S. and R.S.) and 824093 (STRONG- 2020) (R.S.).
Generous computing resources were supplied by the North-German Supercomputing Alliance (HLRN, project bep00072), by the John von Neumann Institute for Computing (NIC) at DESY, Zeuthen and the High Performance Computing Center in Stuttgart (HLRS, PRACE project 5422).

\bibliographystyle{utphys}
\bibliography{skeleton}

\end{document}